\documentclass[pra,showpacs,twocolumn,amssymb]{revtex4}
\newcommand{\be}{\begin{equation}}
\newcommand{\ee}{\end{equation}}
\newcommand{\bea}{\begin{eqnarray}}
\newcommand{\eea}{\end{eqnarray}}
\usepackage{epsfig}
\usepackage{graphicx}
\usepackage{psfrag}

\begin{document}

 \title{Phase Boundary of the Boson Mott Insulator in a Rotating Optical Lattice }

\author{R.~O.~Umucal{\i}lar}
\affiliation{ Department of Physics, Bilkent University, 06800 Ankara, Turkey }%
\author{M.~\"O.~Oktel}
\email{oktel@fen.bilkent.edu.tr}
\affiliation{ Department of Physics, Bilkent University, 06800 Ankara, Turkey }%

\date{\today}

\begin{abstract}

We consider the Bose-Hubbard model in a two dimensional rotating
optical lattice and investigate the consequences of the effective
magnetic field created by rotation. Using a Gutzwiller type
variational wavefunction, we find an analytical expression for the
Mott insulator(MI)-Superfluid(SF) transition boundary in terms of
the maximum eigenvalue of the Hofstadter butterfly. The dependence
of phase boundary on the effective magnetic field is complex,
reflecting the self-similar properties of the single particle
energy spectrum. Finally, we argue that fractional quantum Hall
phases exist close to the MI-SF transition boundaries, including
MI states with particle densities greater than one.

\end{abstract}


\maketitle

Experiments on ultracold atoms in optical lattices opened up a new
avenue to study correlated quantum states \cite{Greiner}. The
versatility of cold atom experiments hold promise for the
experimental realization of many models that were first introduced
for solid-state systems.

\begin{figure}
\includegraphics[scale=0.4]{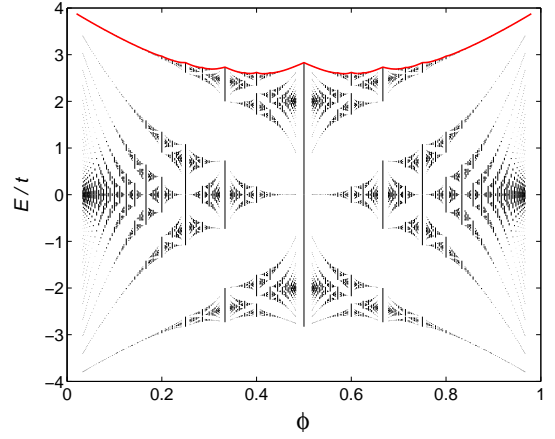}
\caption{(Color online) Maximum energy of the Hofstadter butterfly
$f(\phi)$ for a given $\phi=p/q$. This value is calculated as the
maximum eigenvalue of the matrix
$\mathbb{A}_q=\mathbb{A}_q(k_x=0,k_y=0)$ (Eq.
(\ref{hofsmatrix})).} \label{fig:maxenergy}
\end{figure}

One such model is the study of particles moving in a tight binding
lattice under a magnetic field. When the magnetic flux per
plaquette of the lattice becomes of the order of a flux quantum
$hc/e$, the single particle energy spectrum forms a complicated
self-similar structure, known as the Hofstadter butterfly (Fig.
\ref{fig:maxenergy}) \cite{Hofstadter}. It has not been possible
to reach this regime in ordinary condensed matter experiments due
to the required high magnetic fields. However, the ultracold atom
experiments are extremely flexible and it should be possible to
create required effective magnetic fields in optical lattice
experiments. A conceptually simple way of creating an effective
magnetic field is to rotate the optical lattice, as demonstrated
in a recent experiment \cite{Tung}. Other means of creating
effective magnetic fields  have been discussed by a number of
authors \cite{Jaksch,Sorensen,Mueller,Osterloh,Ruseckas}. Although
the recent demonstration of a rotating optical lattice was done
for a shallow lattice, it should be possible to drive the system
into the Mott insulator (MI) state by increasing the lattice
depth.

In this Letter, we study the Bose-Hubbard model under a magnetic
field. Particularly, we consider a two dimensional square lattice
of spacing $a$ with only nearest neighbor hopping. The magnetic
field (or the effective magnetic field) strength can be expressed
in terms of the dimensionless quantity $\phi$, which is the
magnetic flux quantum per plaquette in the lattice ($a^2H/(hc/e)$,
$H$ being the effective magnetic field). When the Landau gauge
$\overrightarrow{A} = (0,Hx,0)$ is chosen, the Hamiltonian for
this system can be written as
\begin{eqnarray}
H = -t\sum_{\langle i,j\rangle}a_i^\dag a_j e^{i
A_{ij}}+\frac{U}{2}\sum_i \hat{n}_i (\hat{n}_i-1)-\mu \sum_i
\hat{n}_i, \label{hamiltonian1}
\end{eqnarray}
where $a_i$ ($a_i^\dag$) is the bosonic annihilation (creation)
operator at site $i$ and $\hat{n}_i = a_i^\dag a_i$ is the number
operator. The tunnelling strength between nearest neighbor sites
is given as $t$; $U$ is the on-site interaction strength, and
$\mu$ is the chemical potential. Magnetic field affects the
Hamiltonian through $A_{ij}$ which is equal to $\pm 2\pi m \phi$ ,
if $i$ and $j$ have the same $x$ coordinate $ma$ and is $0$
otherwise, while the sign is determined by the hopping direction.

We first review some of the properties of the single particle
spectrum by setting $U = 0$. This problem was first discussed by
Hofstadter \cite{Hofstadter}. The energy spectrum is obtained
through the following difference equation (known as Harper's
equation):
\begin{eqnarray*}
c_{m+1}+c_{m-1} + 2 \cos(2\pi m\phi-k_y)c_m = \frac{E}{t}
c_m,
\end{eqnarray*}
where $c_m$ are the expansion coefficients of the wavefunction,
which has plane wave behavior along $y$ in accordance with the
translational symmetry in this direction. If $\phi$ is a rational
number $p/q$, the wavefunction satisfies the Bloch condition
$c_{m+q}=\exp(i k_x q) c_m$ as a result of the symmetry under
$q$-site translation in the $x$ direction. The allowed energies
are then found as the eigenvalues of the $q \times q$ tridiagonal
matrix:
\begin{eqnarray}
\!\mathbb{A}_{q}(k_x,k_y) \! = \!\! \left(\!%
\begin{array}{ccccc}
  . & \ddots & . & . &  e^{-i k_x q} \\
  \ddots & \ddots & 1 & . & . \\
  . & 1 & 2 \cos(2\pi m\phi-k_y)& 1 & .  \\
  . & . & 1 & \ddots & \ddots \\
  e^{i k_x q}& . & . & \ddots & . \\
\end{array}\!%
\right).\label{hofsmatrix}
\end{eqnarray}
We call the matrix formed by setting $k_x = k_y = 0$ in
(\ref{hofsmatrix}) $\mathbb{A}_{q}$. The maximum eigenvalue of
$\mathbb{A}_{q}$ yields the maximum energy of the system for a
given $\phi$. We define this energy as $f(\phi)$, which is a
continuous but not differentiable function (Fig.
\ref{fig:maxenergy}). To prove that the maximum eigenvalue is
obtained from $\mathbb{A}_{q}$, we investigate the characteristic
equation for the matrix (\ref{hofsmatrix}), which is of the
following form:
\begin{eqnarray}
\bigg(\frac{E}{t}\bigg)^q +\sum_{n=0}^{q-1}a_n
\bigg(\frac{E}{t}\bigg)^n -2cos(k_x q)-2cos(k_y q) = 0.
\label{characteristic}
\end{eqnarray}
Two pairs of $(k_x,k_y)$, namely $(0,0)$ and $(\pi/q,\pi/q)$ are
sufficient to determine the band edges \cite{Thouless}. The
$(0,0)$ pair gives a smaller value for the $k_x$ and $k_y$
dependent terms. Since the $E$ dependent part of
(\ref{characteristic}) increases monotonically after a
sufficiently large $E$, the greatest root is always
obtained from the $(0,0)$ pair.

We now turn to the interacting case with the dimensionless
Hamiltonian:
\begin{eqnarray}
\tilde{H} = -\tilde{t}\sum_{\langle i,j\rangle}a_i^\dag a_j e^{i
A_{ij}}+\frac{1}{2}\sum_i \hat{n}_i (\hat{n}_i-1)-\tilde{\mu
}\sum_i \hat{n}_i \label{hamiltonian},
\end{eqnarray}
where $\tilde{t}=t/U$ and $\tilde{\mu}=\mu/U$ are the scaled
hopping strength and chemical potential.

When the hopping term is dominant $\tilde{t} \gg 1$, one expects
the system to be in a SF state, while in the opposite limit of
strong interactions $\tilde{t} \ll 1$, the system should go into
the MI state. In this Letter, we investigate the transition
boundary between these two phases, and how this boundary is
affected by the external magnetic field. The effect of the
magnetic field on the transition boundary has been previously
explored by strong coupling expansion for small magnetic fields by
Niemeyer et. al. \cite{Niemeyer}, and numerically within
mean-field theory by Oktel et. al. \cite{Oktel}. Here we use a
variational approach to provide an analytical expression for the
transition boundary.

We use a site dependent Gutzwiller ansatz to describe the system
\cite{Rokhsar}. For the Bose-Hubbard  model without magnetic
field, this ansatz (and equivalent mean-field theory
\cite{Sheshadri,van Oosten}) gives an accurate description of the
phase diagram. We introduce the variational wave function at each
site $l$,
 \bea
 |G\rangle_l = \Delta_l|n_0-1\rangle_l+|n_0\rangle_l+\Delta_l^{\prime}|n_0+1\rangle_l
\label{variational}.
\eea
Since we investigate the behavior in the
vicinity of the transition region, we consider small variations
around the perfect MI state with exactly $n_0$
particles per site, allowing for only one less or one more
particle in a site. The variational parameters $\Delta_l$ and
$\Delta_l^{\prime}$ are assumed to be real, as complex $\Delta$
values can only increase the energy of the variational state.
Total wavefunction is the direct product of these site
wavefunctions $|\Psi\rangle = \prod_i^N |G\rangle_i$. Within the
selected gauge, the magnetic Hamiltonian has translational
invariance in the $y$ direction. The translational invariance in
the $x$ direction is broken by the magnetic field, but can be
restored to a certain degree if the flux per plaquette is a
rational number. Thus, taking $ \phi = p/q$ where $p$ and $q$ are
relatively prime integers, the Hamiltonian is invariant under
translation by $q$ sites in the $x$ direction. This periodicity
simplifies the calculation of the expectation value of the energy
when we work with a supercell of $1 \times q$ sites. Total
wavefunction for such a supercell is $|\Psi\rangle_s =
\prod_{l=0}^{q-1} |G\rangle_l$. The expected value of the energy
can then be written as follows
\begin{eqnarray}
\frac{\langle\Psi|\tilde{H}|\Psi\rangle}{\langle\Psi|\Psi\rangle}
= N_s
\frac{_s\langle\Psi|\tilde{H}|\Psi\rangle_s}{_s\langle\Psi|\Psi\rangle_s}
\equiv N_s\varepsilon \label{energy},
\end{eqnarray}
where $N_s$ is the number of supercells.

Keeping terms up to second order in the variational parameters
$\Delta$, the energy of a supercell is calculated as
\begin{widetext}
\begin{eqnarray}
\varepsilon &=&
\sum_{l=0}^{q-1}\bigg[-2\tilde{t}\Big\{n_0\Delta_{l}\Delta_{l+1}+\sqrt{n_0(n_0+1)}\Delta_{l}\Delta_{l+1}^{\prime}+\sqrt{n_0(n_0+1)}\Delta_{l+1}\Delta_{l}^{\prime}
+(n_0+1)\Delta_{l}^{\prime}\Delta_{l+1}^{\prime} \nonumber \\ &
&+\cos(2\pi\frac{p}{q}l)\big[ n_0\Delta_{l}^{2}+2\sqrt{n_0(n_0+1)}\Delta_{l}\Delta_{l}^{\prime}+(n_0+1)(\Delta_{l}^{\prime})^2\big] \Big\} \nonumber \\
&
&+\frac{1}{2}\big[2(1-n_0)\Delta_{l}^{2}+2n_0(\Delta_{l}^{\prime})^2+n_0(n_0-1)\big]+\tilde{\mu}\big[\Delta_{l}^{2}-(\Delta_{l}^{\prime})^2-n_0\big]
\bigg]. \label{varepsilon}
\end{eqnarray}
\end{widetext}

\begin{figure}
\includegraphics[scale=.9]{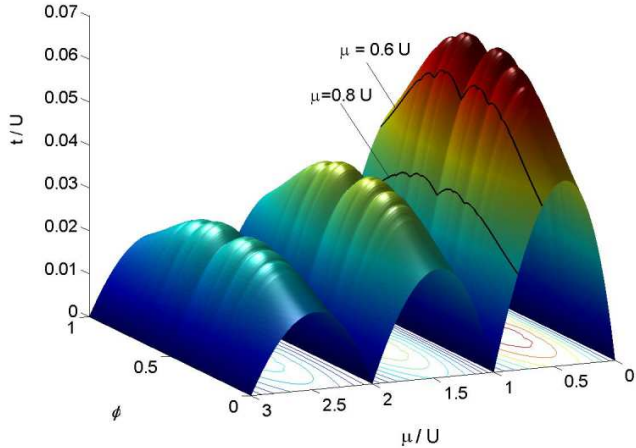}
\caption{(Color online) The boundary of the Mott insulator phase for the first
three Mott lobes. The figure is periodic in $\phi$. Magnetic field
increases the critical value for $t/U$, as expected, however
this increase is not monotonic. Transition boundary for two
different values of $\mu/U$ are marked to display the
complex structure of the surface.} \label{fig:hills}
\end{figure}

If the system favors to be in the Mott insulator state, the total
energy of the system should be a minimum where all the variational
parameters vanish. Thus, we can find the phase boundary as the
point where the total energy ceases to be a local minimum in
$\Delta$. As a result, we demand that the matrix composed of the
second derivatives of $\varepsilon$ with respect to the parameters
($\partial^2 \varepsilon / \partial \Delta_i \partial \Delta_j$,
$\partial^2 \varepsilon / \partial \Delta_i \partial
\Delta^{\prime}_j$, $\partial^2 \varepsilon / \partial
\Delta_i^{\prime}
\partial \Delta^{\prime}_j $) be positive definite, i.e. all
eigenvalues be positive. This matrix can be written compactly as:
\begin{eqnarray*}
\mathbb{F} &=& -2\tilde{t}\left(%
\begin{array}{cc}
  n_0\mathbb{A}_{q} & \sqrt{n_0(n_0+1)}\mathbb{A}_{q} \\
  \sqrt{n_0(n_0+1)}\mathbb{A}_{q} & (n_0+1)\mathbb{A}_{q} \\
\end{array}%
\right)\\ &&+\left(%
\begin{array}{cc}
  2(1-n_0+\tilde{\mu})\mathbb{I}_{q} & 0 \\
  0 & 2(n_0-\tilde{\mu})\mathbb{I}_{q} \\
\end{array}%
\right),
\end{eqnarray*}
where $\mathbb{I}_{q}$ is the $q \times q$ identity matrix, and
$\mathbb{A}_{q}$ was introduced before (Eq. \ref{hofsmatrix}).

If we denote the eigenvalues and eigenvectors of $\mathbb{A}_{q}$
by $\lambda_{\nu}$ and $\overrightarrow{\nu}$, and those of
$\mathbb{F}$ by $\lambda_u$ and $\overrightarrow{u}$, all
$\lambda_u$ can be expressed in terms of $\lambda_{\nu}$ by taking
\begin{eqnarray*}
\overrightarrow{u} = \left(%
\begin{array}{c}
  a \overrightarrow{\nu}\\
  b \overrightarrow{\nu} \\
\end{array}%
\right),
\end{eqnarray*}
due to the special block form of $\mathbb{F}$. Then $\lambda_u$
are obtained as:
\begin{widetext}
\begin{eqnarray*}
\lambda_u^{\mp} = 1-(1+2n_0)\tilde{t} \lambda_{\nu} \mp
\sqrt{\big[(1+2n_0)\tilde{t}
\lambda_{\nu}-1\big]^2-4\big\{(n_0-\tilde{\mu})[1-(n_0-\tilde{\mu})]-\tilde{t}(1+\tilde{\mu})\lambda_{\nu}\big\}}.
\end{eqnarray*}
\end{widetext}
The positive definiteness of $\mathbb{F}$ leads us to take
$\lambda_u^{-}$ and set it to 0 in order to determine the critical
$\tilde{t}$ value above which the perfect insulator state is
destroyed. We find the boundary of the $n_0^{\rm{th}}$ Mott lobe to be:
\be \tilde{t}_c =
\frac{(n_0-\tilde{\mu})[1-(n_0-\tilde{\mu})]}{(1+\tilde{\mu})f(\phi)}
\label{tcritical}, \ee where $n_0-1 \leq \tilde{\mu} \leq n_0$.
This boundary is plotted in Fig. \ref{fig:hills} for the first
three Mott lobes. At $\phi = 0$, this formula reproduces the
critical $\tilde{t}$ value found in \cite{Sheshadri,van Oosten}.
Increasing magnetic field increases the critical hopping strength
$\tilde{t}_c$, however this increase is not monotonic. The
complicated structure of the single particle problem is reflected
in the transition boundary. Equation (\ref{tcritical}) is in
excellent agreement with the numerical mean-field work
\cite{Oktel}.

We can comment on the accuracy of our variational approach. Our
result is exact within mean-field theory. At zero magnetic field
the mean-field result for the transition boundary is close to
accurate Monte Carlo calculations \cite{Krauth}, but it is not
guaranteed that the mean-field description of the system would be
valid under magnetic field. Our variational wavefunction (and
mean-field theory) disregards the correlations between
fluctuations above the insulating state. Such correlations would
be expected to wash out the fine structure of the transition
boundary (Fig. 2). Nevertheless, one can expect a number of
features of the mean-field boundary to survive for the real
system. The linear increase of the transition point for small
magnetic fields, periodicity of the system with $\phi$, and the
central dip near $\phi=0.5$ should be qualitatively correct.

There is however one important way that the fluctuations around
the Mott insulating state can become correlated. The Hamiltonian
(\ref{hamiltonian}) supports bosonic fractional quantum Hall (FQH)
states as discussed in a number of recent papers
\cite{Sorensen,Bhat,Palmer}. So far, such FQH states have been
assumed to appear only in the region of low density where the
number of particles per site is less then one. Here, we argue that
states similar to bosonic FQH states should be present near the
MI boundaries, even at higher densities.

It is instructive to think about the behavior of the Hamiltonian
for constant particle density by disregarding the last term. Let
us assume that the particle density is equal to $n=n_0+\epsilon$,
where $n_0$ is an integer and $\epsilon << 1$ is the decimal part
of the density. With such incommensurate particle number, the
system never goes into the MI state, but will always have a
superfluid density. The chemical potential for this state, plotted
on the $\tilde{\mu},\tilde{t}$ plane, traces the outline of the
Mott lobe as the interaction is increased (Fig. 3). However, if we
think of the same system under a magnetic field that is
commensurate with the excess particle density, another possibility
presents itself. Specifically, considering a magnetic field so
that $\phi= 2 \epsilon$, it is possible for $n_0$ particles to
form a MI state that is coexisting with a $\nu=1/2$ Bosonic
Laughlin state of the remaining $\epsilon$ particles. At high
enough interaction, such a state would be preferable to a
superfluid state as it avoids any interaction between the
``excess" particles.
\begin{figure}
\includegraphics[scale=0.35]{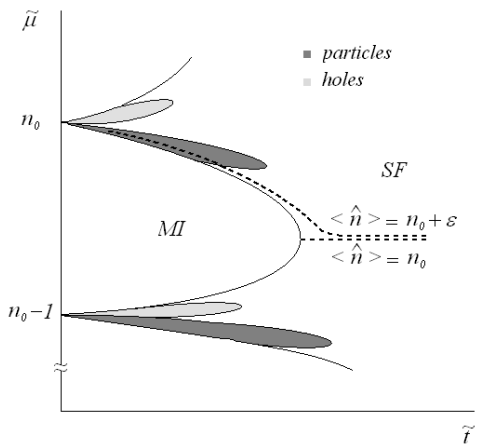}
\caption{Schematic phase diagram near the
$n_0^{\rm{th}}$ Mott lobe. Dotted lines show the chemical potential as a function of
hopping strength for systems with constant density $\langle \hat{n} \rangle = n_0$ and $\langle \hat{n} \rangle = n_0 + \varepsilon$.
FQH phases of ``excess" particles, or holes are shown as the shaded regions.} \label{fig:qhs}
\end{figure}
The wavefunction of such a state can be obtained by symmetrizing
the product of the Mott insulator state for $n_0$ bosons with the
$\nu=1/2$ Bosonic Laughlin state for $\epsilon$ particles. In
general, separating the many particle wavefunction into two parts
and arguing that the overall properties can be deduced by thinking
about the individual parts is not correct, as symmetrization may
change the character of both parts considerably. In this case,
however we can safely regard  the excess particles as
forming a correlated state above the Mott insulator, due to the
full translational invariance of the MI state. One can write
down an effective Hamiltonian for the excess particles. To the
zeroth order, the change in the
effective Hamiltonian would be just to replace $t$ by $(n_0+1)t$,
due to bosonic enhancement of the hopping. There will be higher
order corrections to $t$ and new non-contact interaction terms
between the excess particles due to fluctuations in the MI state.
Such terms will be of higher order in (t/U), and can be neglected
in the strongly interacting limit. One can also argue that as both
the MI state and the Bosonic Laughlin state are gapped, it would
not be energetically favorable to exchange particles between the
two parts of the wavefunction. Similarly, one can argue that the
overall state would be gapped in the strongly interacting limit.

Treating such a state as a variational state, the energy
difference from the MI state can be written as \be \Delta E =
\left( U n_0 - \mu - t (n_0 + 1) f(\phi) \right) \epsilon. \ee To
first order in $t/U$, the term in parenthesis is the energy needed
to put one extra particle on to the Mott insulator. Thus, when it
is favorable to put one extra particle on to the Mott state, it
would be favorable to put more particles (up to $\epsilon$ per
site) and organize them into a FQH state. One can then expect the
correlated state to exist within a band above the MI lobe (see
Fig. \ref{fig:qhs}). The same argument can be advanced for holes
in a MI state, creating a FQH of holes below the Mott insulator.
Experimentally these states would have distinct signatures
appearing as extra steps near the MI steps of the Ziggurat
structure of the trapped MI. Detailed properties of these
correlated states, as well as other correlated states near the
transition boundary will be investigated elsewhere
\cite{Umucalilar}.

In conclusion,  we studied the phase boundary of the  MI state of
bosons in a rotating optical lattice. Using a Gutzwiller ansatz,
we gave an analytical expression for the phase boundary in terms
of the maximum energy of the Hofstadter butterfly. We finally
argued that analogues of FQH states will be found close to the
MI-SF transition boundary including MI states with particle
densities greater than one.

\begin{acknowledgements}
R.O.U. is supported by TUBITAK. M.\"O.O. wishes to thank B. Tanatar, M. Ni\c{t}\u{a} and Qi Zhou
for useful discussions. This work was partially supported by a TUBA-GEBIP grant and TUBITAK-KARIYER
grant No. 104T165.
\end{acknowledgements}

\end{document}